\documentclass[twocolumn,aps,prb,10pt,floatfix,superscriptaddress,longbibliography]{revtex4-2}
\usepackage{graphicx}
\usepackage{dcolumn}
\usepackage{bm}
\usepackage{amsmath}
\usepackage{amssymb}
\usepackage{siunitx}
\usepackage{color}
\usepackage{hyperref}
\usepackage{physics}
\usepackage{float}

\begin{document}

\preprint{APS/123-QED}

\title{Creation and motion of antiferromagnetic skyrmions by edge manipulation}

\author{Aleksey Berg}
\affiliation{Department of Physics, University of Hamburg, 20355 Hamburg, Germany}
\author{Tim Matthies}
\affiliation{Department of Physics, University of Hamburg, 20355 Hamburg, Germany}
\author{Roland Wiesendanger}
\affiliation{Department of Physics, University of Hamburg, 20355 Hamburg, Germany}
\author{Elena Y. Vedmedenko}
\affiliation{Department of Physics, University of Hamburg, 20355 Hamburg, Germany}

\date{\today}

\begin{abstract}
Magnetic racetrack architectures that use topological magnetic particles to store information are one of the most promising concepts for future storage applications. Antiferromagnetic racetracks are particularly appealing as they are not susceptible to external magnetic fields. State-of-the-art racetracks use magnetic fields, spin-transfer and spin-orbit torques caused by electric currents to move the bits across the entire circuit. However, the application of currents in many antiferromagnetic racetracks is limited because many of them are insulating. Recently, however, a concept for ferromagnetic racetrack memories that are free of global driving forces has been proposed. It has been demonstrated that various topological entities can be generated and transported over long distances solely through local magnetization rotation at the sample boundaries, independent of global driving forces. Here, we demonstrate that the local rotation of magnetization at the boundary of an antiferromagnetic sample can be exploited in racetracks to efficiently generate and transmit antiferromagnetic skyrmions. Additionally, we demonstrate that local switching of staggered magnetization at the edge of an antiferromagnetic racetrack can be even more successful than the  rotational procedure. A comparison of ferromagnetic and antiferromagnetic processing of skyrmionic bits, together with energy considerations, shows that this procedure is fairly efficient in antiferromagnets. 
\end{abstract}
\maketitle

\section{\label{sec:int}Introduction}

Antiferromagnets (AFM) are materials without macroscopic magnetization, but with spatially dependent local alignment of  magnetic moments. These materials are at the heart of the antiferromagnetic spintronics - the research field studying the possibilities to control the magnetization texture without being influenced by external magnetic fields \cite{Zhang:SR2016,Jungwirth:Nnano2016,Din:npjSpin2024}. One of the biggest challenges for the antiferromagnetic spintronics is that the spins in antiferromagnetic materials are difficult to manipulate, because they are not susceptible to external magnetic fields and very often are insulating; that is, the application of electric currents is also limited. Finding a way to control antiferromagnetic textures reliably and precisely is crucial to create memory devices.

Currently, AFM can be manipulated using spin-transfer torques in conducting materials \cite{Wadley:Science2016} or by utilizing the spin Hall effect in insulators  \cite{Shi:NatElectron2020}. These switching mechanisms have been applied to nanoscopic AF cells, similar to magnetic random-access memories (MRAMs), with the aim of rotating the Néel vector between two perpendicular directions \cite{Zhang:SR2016,Jin:APL2016}. Apart from MRAMs and magnetic cells, another promising concept for magnetic memory is magnetic racetracks. In this concept, bits of information correspond to topological, magnetic, particle-like textures (such as skyrmions~\cite{fert2013skyrmions,gobel2019electrical}, domain walls and bimerons), which can be moved along a long magnetic stripe, called a 'racetrack', by currents, field gradients or spin waves  \cite{Parkin:NatNano2015}. Numerous memory and computational concepts have been proposed for magnetic skyrmions~\cite{finocchio2016magnetic}, e.g., logic gates~\cite{zhang2015logic} and skyrmion-based nano-oscillators~\cite{zhang2015Skyrosc}. Skyrmions can navigate racetracks with not ideal conditions, like notches~\cite{Morshed2022Notched} or pinning sites~\cite{jiang2017direct,woo2016observation}. Proposals have been made to use the spin Hall effect for topological particles in the case of natural AFM \cite{Xia:JphysD2017} or to propel topological objects using spin-orbit torques in synthetic antiferromagnetic materials (SAFs), which consist of two ferromagnetic layers separated by a non-magnetic spacer and coupled antiferromagnetically. The latter concept has been realized experimentally \cite{Barker:JPhysD2023}. In all the above cases, electric currents or field gradients must be applied to the entire magnetic structure, or 'racetrack'.

We investigate antiferromagnetic skyrmions which are more recent and less studied topological objects within the context of racetrack memory \cite{Zhang:SR2016,Baker:PRL2016}. An advantage of antiferromagnetic skyrmions is the vanishing skyrmion Hall angle or Magnus force \cite{Zhang:SR2016,Baker:PRL2016}. They can be moved by electric currents \cite{Zhang:SR2016,Xia:JphysD2017,Baker:PRL2016,Aldarawsheh:Spint2024}. Various methods have been proposed for insulating AFMs, such as anisotropy gradients \cite{Shen:PRB2018}, spin waves \cite{Lau:PRB2024}, and sublattice displacements \cite{Lau:PRB2025}.

We have recently demonstrated that ferromagnetic topological magnetic structures can be created and manipulated without the use of global fields or currents. This can be achieved by applying time-dependent boundary conditions to one-, two-, and three-dimensional classical and quantum mechanical racetracks \cite{Vedmedenko:PRL2014,Schaeffer:SciRep2020,Siegl:PhysRevB2022,Siegl:PhysRevR2022}. These results, which are based on atomistic calculations, have been strongly corroborated by large-scale micromagnetic simulations \cite{Dzem:SciRep2015,McGrath:PhysRevB2025} and experiments \cite{Huang:SciRep2022}. 

Here, we explore the creation and processing of antiferromagnetic skyrmions by manipulating the boundary magnetization of a nanoscale racetrack exhibiting antiferromagnetic exchange interactions, Dzyaloshinskii–Moriya interactions (DMI), and uniaxial magnetic anisotropy. First, we present an adiabatic boundary rotation scheme for creating antiferromagnetic skyrmions and compare it with the scheme used for ferromagnetic skyrmions. Secondly, we demonstrate that the rotation of the boundary can be effectively replaced by simply switching the boundary element of the racetrack. Thirdly, we analyze how stable and effective the proposed procedure is.

\section{\label{sec:theory}Model}

We consider a square lattice in the form of a rectangular racetrack with $\tilde{N}_x \times \tilde{N}_y$ lattice sites. Unless specified otherwise, we choose $\tilde{N}_x = 120, \tilde{N}_y = 28$ with open boundary conditions. 
The atomistic Hamiltonian for our investigation reads 
\begin{equation}
\begin{split}
H = &-J \sum_{<ij>} \mathbf{S}_i \cdot \mathbf{S}_j -\sum_{<ij>} \mathbf{D}_{ij} \cdot (\mathbf{S}_i \times \mathbf{S}_j)  \\ &- K \sum_i \left( \mathbf{S}_i^z \right)^2 -\mu_s \sum_i \left( \mathbf{B}_{\text{edge},i} \cdot \mathbf{S}_i \right),
\end{split}
\label{eq:hamiltonian}
\end{equation}
with the nearest neighbor exchange interaction constant $J$, the DMI vectors $\mathbf{D}_{ij}$, and the magnetic anisotropy constant $K$. Additionally, we consider a Zeeman-like interaction term for an applied edge field $\mathbf{B}_{\text{edge},i}$. When considering an antiferromagnetic system, the edge field changes sign for the two different sublattices. Here, the strength of the magnetic moments $\mu_s$ is assumed to be equal to $3\mu_B$, like in~\cite{Spethmann:CommPhys2022}.
The magnetic moments $\mathbf{S}_i$ are assumed to be normalized, i.e., $|\mathbf{S}_i| = 1$. The last sum in Eq. \eqref{eq:hamiltonian} runs over all lattice sites, while the first two sums run over all nearest neighbors.
More specifically, the interaction parameters were set to $J = \pm\SI{11.6}{meV}$, $|\mathbf{D}_{ij}| = \SI{3.17}{meV}$ and $K = \SI{1.15}{meV}$. These values fall within the range observed in Pd/Fe/Ir(111), Mn/W(001), and Fe/W(001) monolayers \cite{Bode:NatMat2006,Bruening:PhysRevB2022,Romming:PhysRevLett2015}. The sign of $J$ determines if the system is a ferromagnet (FM), $J>0$, or an AFM with $J<0$.
The time evolution of the system is described by the Landau-Lifshitz-Gilbert equation (LLG)
\begin{equation}
\frac{\partial \mathbf{S}_i}{\partial t} = -\frac{\gamma}{1 + \alpha^2} (\mathbf{S} \times \mathbf{B}_{\text{eff},i} + \alpha \mathbf{S} \times (\mathbf{S} \times \mathbf{B}_{\text{eff},i})),
\label{eq:LLG}
\end{equation} 
with the gyromagnetic ratio $\gamma$ and the damping constant $\alpha = 0.1$. The effective magnetic field is defined as $B_\text{eff} \equiv \frac{1}{\mu_s} \frac{\partial H}{\partial \mathbf{S}}.$
By assuming that a skyrmion is a rigid structure on the racetrack, we can apply the Thiele equation \cite{Thiele:PRL1973} to describe the skyrmion dynamics analytically
\begin{equation}
\mathbf{G} \times \mathbf{v} - \alpha \mathcal{D} \mathbf{v} + \mathbf{F} = 0.
\label{eq:thiele_general}
\end{equation}
Here, $\mathbf{v}$ is the drift velocity of the skyrmion and $\mathbf{F}$ the sum of all forces acting on the skyrmion.  Furthermore, Eq. \eqref{eq:thiele_general} contains the gyro-coupling vector $\mathbf{G}$ with vector components $\text{G}_k = \frac{1}{4\pi} \epsilon_{ijk} \int \mathbf{S} \cdot (\partial_i \mathbf{S} \times \partial_j \mathbf{S}) \text{ } \dd^2 r$ and the dissipative tensor $\mathcal{D}_{ij} = \frac{1}{2\pi} \int \partial_i \mathbf{S} \cdot \partial_j \mathbf{S} \text{ } \dd^2 r$.
We can solve Thiele's equation for the drift velocity, which yields the following result
\begin{equation}
\mathbf{v} = \frac{1}{G^2 + \alpha^2 D^2} (\mathbf{G} \times \mathbf{F} + \alpha \mathcal{D} \mathbf{F}).
\label{eq:thiele_velocity}
\end{equation}
Here, $G = |\mathbf{G}|$ and $D = |\mathcal{D}|$. For ferromagnetic skyrmions, we define the skyrmion number $N_\text{T}$ via $N_\text{T} = \text{G}_z$. For antiferromagnetic ones, we flip the spins in one of the two sublattices and then calculate $\text{G}_z$ with the method presented in~\cite{berg1981TopoCharg}. In conclusion, the skyrmion drift velocity $\mathbf{v}$ is determined by two different contributions of force terms. The latter is the drag force, which determines the skyrmions' velocity component in the direction of $\mathbf{F}$. The first term is the Magnus force, resulting in a non-zero velocity component perpendicular to the sum of all forces $\mathbf{F}$ if $G \neq 0$.
\begin{figure}
\includegraphics[width=\linewidth]{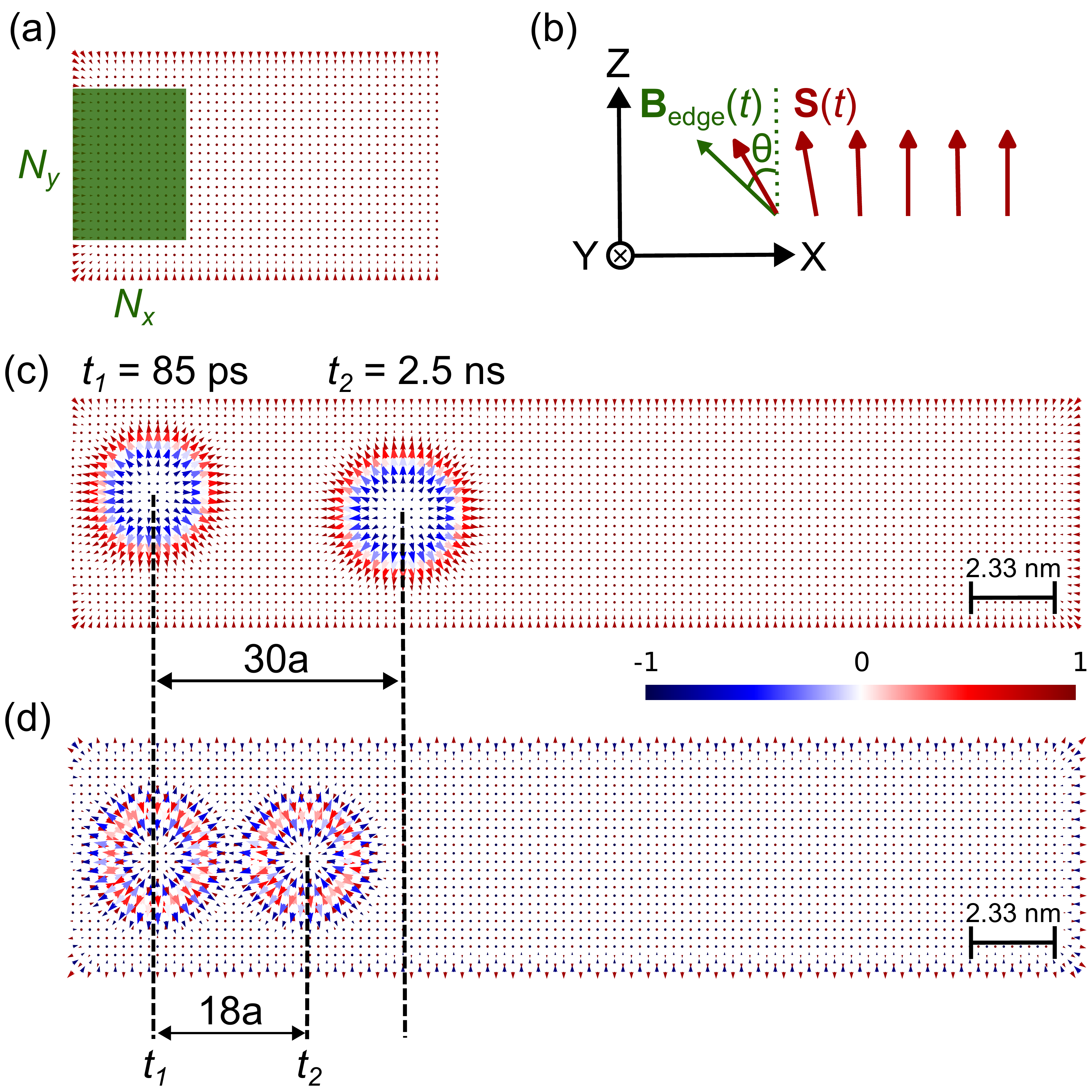}
\caption{(a) Section of a ferromagnetic rectangular racetrack. The green rectangle indicates the region which the edge field is applied to. $N_x$ is the number of affected rows along the $x$ axis while $N_y$ stands for the number of affected columns along the $y$ axis. (b) Side view of the ferromagnetic racetrack at the left edge to schematically illustrate the rotation of applied local field (green arrow) and its action on the magnetization (red arrows). Here, the edge field rotates in the $xz$ plane or switches periodically along a previously defined axis. The axis is defined by the angle $\theta$ with respect to the $z$ axis. (c) shows the position of a ferromagnetic skyrmion and (d) antiferromagnetic one at given times $t$  after a single $2\pi$ rotation, where $t=0$ is the moment the rotating edge field is applied to the system. After the single $2\pi$ rotation the system is allowed to relax. The lattice constant reads $\text{a} = \SI{0.233}{nm}$ and the frequency of the rotation is $\nu=\SI{10}{GHz}$ for both. For (c) and (d), $N_x = 5$ and $N_y = 14$.}
\label{fig:edge}

\end{figure}
\section{\label{sec:results}Comparison of local creation of FM and AFM skyrmions by edge rotation}

In order to store information using skyrmions in a racetrack, these magnetic objects must be created, deleted and propelled in a controlled manner. As demonstrated in \cite{Schaeffer:SciRep2020,Siegl:PhysRevB2022} ferromagnetic skyrmions can be generated and manipulated by applying a smoothly rotating local magnetic field to the magnetic moments at the edge of a racetrack. 
To apply a rotating edge field to an antiferromagnetic racetrack, the track can be decorated using ferromagnetic material, as shown in \cite{Spethmann:CommPhys2022}. Assuming $\mu_{\rm s} \approx \mu_{\rm B}$ and that the exchange interaction between FM and AFM $J^{\rm AFM-FM}$ ranges from 1 to 10 meV, the effective field $\mathbf{B}_{\rm eff,i}$ is of the order of $\SIrange{16}{160}{T}$, guided by the reported interactions of CoFeB/IrMn~\cite{Zhang2022CoFeBonIrMn}. The ferromagnetic patch has a non-zero magnetic moment and can be rotated via an external magnetic field if the magnetic anisotropy is sufficiently small. Possible candidates include soft magnetic materials such as permalloy~\cite{Rodrigues2018permalloy,Klaui2023Permalloy} and CoFeB~\cite{bilzer2006CoFeBSoft,xie2020magnetocrystalline}. Fig.~\ref{fig:edge} (a-b) show the top and the side views of such a racetrack and an edge region (green rectangle) affected by a local time-depending magnetic field $\mathbf{B}_{\rm edge}(t)$. Here, we discuss two scenarios: (i) the edge field rotates smoothly in the $xz$ plane at a constant frequency $\nu$ according to the expression $\mathbf{B}_{\rm edge}(t)=(\sin{(2\pi\nu t)},0,\cos{(2\pi\nu t)})$; (ii) the edge field switches at a constant frequency $\nu$ between two stable antiparallel orientations in the $xz$ plane making an angle of $\theta$ with respect to the $z$ axis. We specify the first scenario as 'rotating field' and the second as 'switching field'. Rather than applying the edge field only to the left edge of the racetrack, we analyze the impact that the amplitude of the edge field, as well as the additional rotation of neighboring magnetic and edge rows, has on the creation process in ferromagnetic and antiferromagnetic cases.

\begin{figure}[hbt!]
\centering
\includegraphics[width=0.8\linewidth]{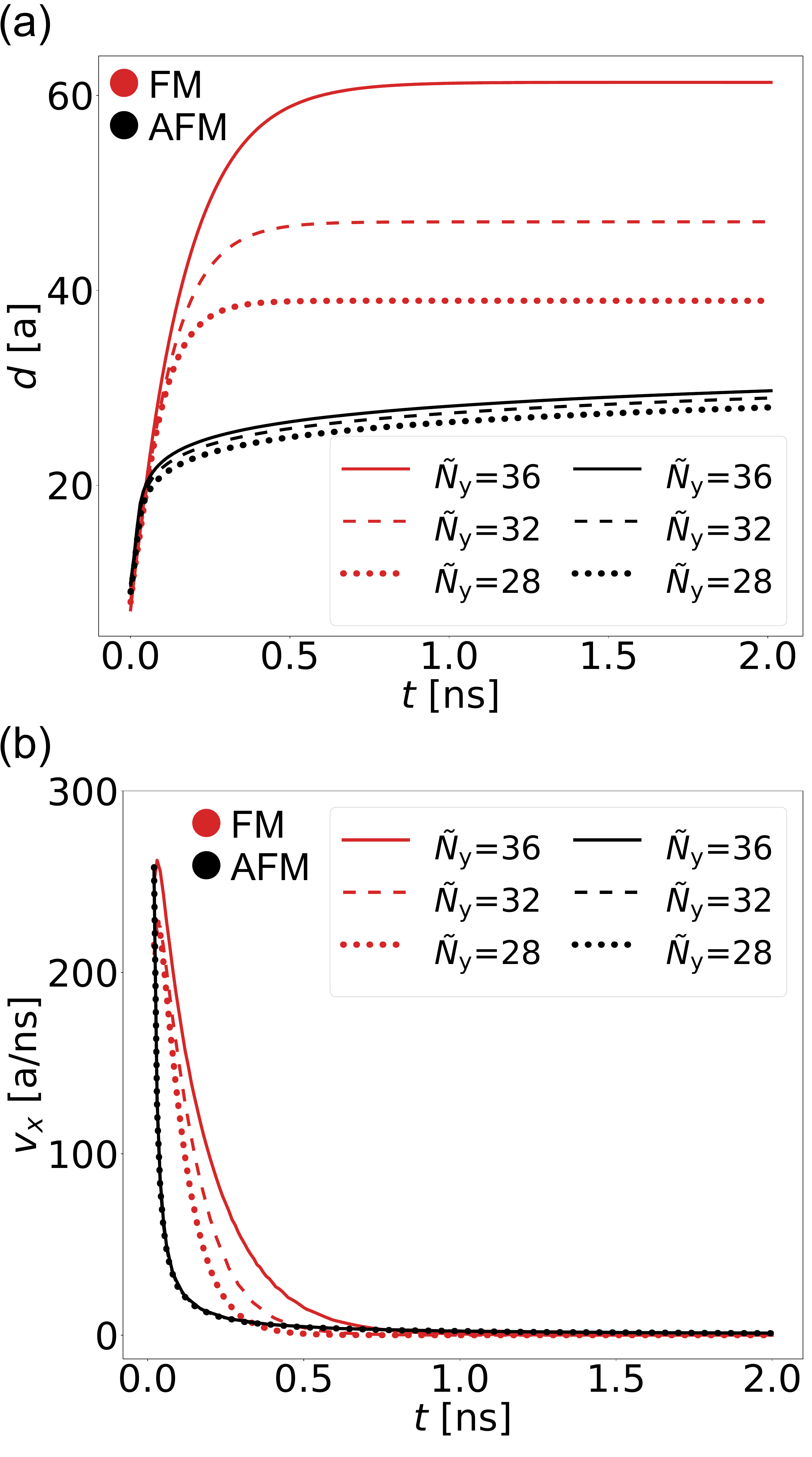}
\caption{(a) Time-dependent distance $d$ of a skyrmion measured after a $2\pi$ edge field rotation at the left edge of a $120 \times \tilde{N}_y$ racetrack. Here, $t=0$ corresponds to the time value $t_1 = \SI{85}{ps}$ in Fig.~\ref{fig:edge}. After the single $2\pi$ rotation the system is left to relax. (b) Time-dependent x-component of the skyrmion drift velocity $v_x$. The red lines correspond to a FM and the black lines to an AFM. The number of rotated edge rows is $N_x = 5$ and $N_y = 14$. The edge field amplitude is given by $B_\text{edge} = \SI{10}{T}$.}
\label{fig:trajectory}
\end{figure}

To compare the edge creation of ferromagnetic and antiferromagnetic skyrmions by a rotating field, we initially apply a local effective edge field, which rotates once within $\SI{100}{ps}$ (at a frequency of $\SI{10}{GHz}$) across the entire edge, before being deactivated in a manner consistent with a previous study \cite{Schaeffer:SciRep2020}. Both the ferromagnetic and antiferromagnetic skyrmions were successfully inscribed into the racetrack. Fig.~\ref{fig:edge}(c) and (d) show the positions of the skyrmions created at two specific times, $t_1$ and $t_2$, following one full rotation of the $\mathbf{B}_{\rm edge}$ field followed by deactivation of the rotation. We used the damping parameter of $\alpha=0.1$. As can be seen from these images, the trajectory of the ferromagnetic skyrmion deviates from the center of the racetrack due to edge repulsion and the resulting Magnus force. In contrast, the antiferromagnetic skyrmion propagates along a straight trajectory because the Magnus force is perfectly compensated for by the two staggered sublattices. The lateral displacement of the antiferromagnetic skyrmion is somewhat shorter than that of its ferromagnetic counterpart. 

This tendency remains true for all geometries of the rotated area of the racetrack, as can be seen in Fig.~\ref{fig:trajectory}. For simplicity, we fixed the width of the rotated area in the $x$ direction to five atomic rows ($N_x=5$) and varied the racetrack width in the $y$ direction between 20 and 40 atomic distanceswhile keeping $N_y = 14$ constant. In all cases, the lateral displacement $d$ of ferromagnetic skyrmions (red lines) was larger than that of antiferromagnetic skyrmions (black lines). Increasing $\tilde{N}_y$ leads to an increase in displacement. Not only was the distance travelled by the antiferromagnetic skyrmions smaller, but they were also more inert than the ferromagnetic ones. While ferromagnetic skyrmions reach maximal displacement within several hundred picoseconds, antiferromagnetic skyrmions stop on the scale of nanoseconds. That is, the propagation velocity of the antiferromagnetic skyrmions is lower. The main reason a FM skyrmion takes a longer travelling distance is its repulsive interaction with the long boundaries of the racetrack. This repulsion enhances the driving force due to the edge rotation as can be seen from the Thiele equation (\ref{eq:thiele_general},\ref{eq:thiele_velocity}). In the AFM case, however, this side repulsion is completely compensated.

It should be noted that the observed plateaus in the FM skyrmion trajectories are due to numerical limitations. In theory, a skyrmion would always relax into the center of the racetrack given at $d=60$a. Resolving this issue would require solving the LLG equation for infinitely small time steps $\Delta t \rightarrow 0$ and for longer times.

\begin{figure}[hbt!]
\includegraphics[width=0.95\linewidth]{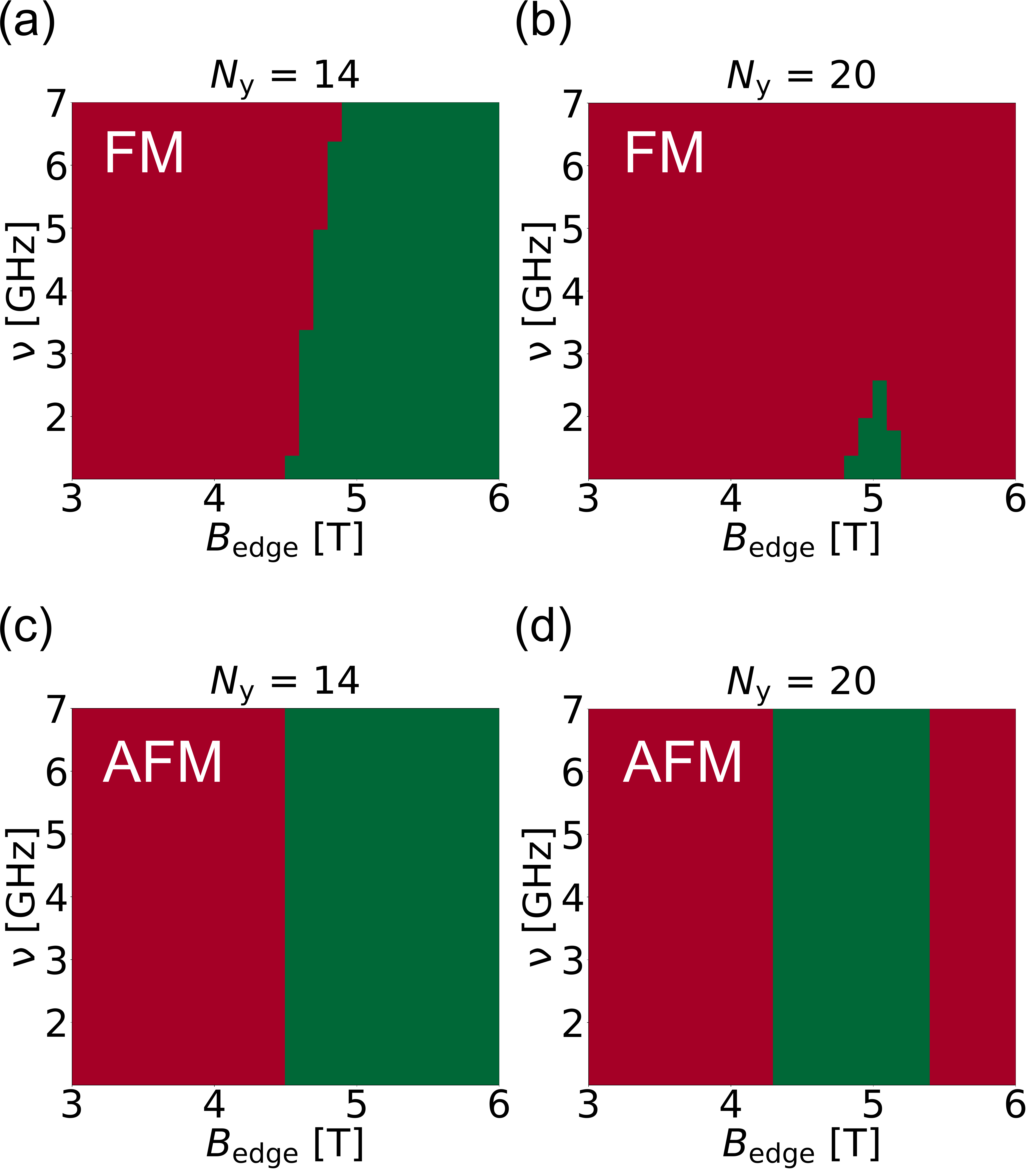}
\caption{Skyrmion creation diagrams with rotating edge fields for different pairs of edge field amplitudes $B_\text{edge}$ and edge field frequencies $\nu$ in a FM, (a)/(b), and in an AFM, (c)/(d). The number of rotated edge rows is set to $N_x=5$. The step sizes are $\Delta B_\text{edge} = \SI{0.1}{T}$, $\Delta \nu = \SI{0.2}{GHz}$. Green areas indicate a skyrmion number $|N_\text{T}| = 1$ after one $2\pi$ rotation and $\SI{200}{ps}$ of relaxation. Red areas refer to $N_\text{T} = 0$.}
\label{fig:creation_diagrams}
\end{figure}
While the amplitude of the rotating field was chosen to be very high ($B_\text{edge} = \SI{10}{T}$) in the previous discussion, it can be significantly reduced by either increasing the area over which the field rotates or decreasing the frequency of the edge rotation, as shown in the diagrams in Fig.~\ref{fig:creation_diagrams}. Additionally, $B_\text{edge}$ may correspond to an effective field stemming from exchange interactions, rather than the external magnetic field. The diagrams of Fig.~\ref{fig:creation_diagrams} illustrate the skyrmion number of the racetrack following one complete edge rotation at a given frequency $\nu$, and a given strength of the edge field $B_\text{edge}$. The green areas correspond to a skyrmion number of unity, $|N_{\text{T}}| = 1$, while the red areas correspond to $|N_{\text{T}}| = 0$. Direct comparison of Figs.~\ref{fig:creation_diagrams} (b) and (d) shows that the area of stable skyrmion creation is larger for AFMs compared to FMs for the same set of parameters.

\begin{figure}[hbt!]
\includegraphics[width=\linewidth]{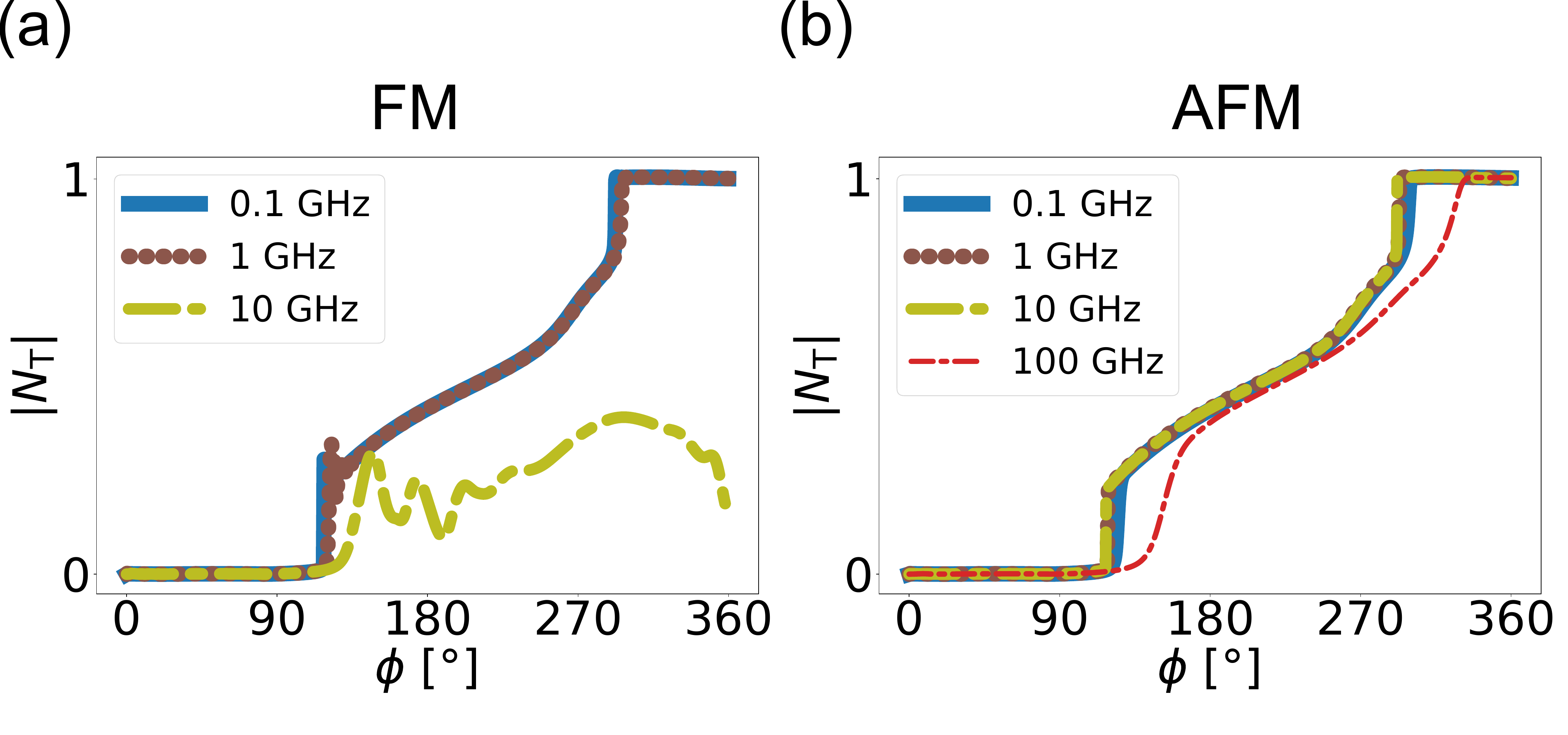}
\caption{Time-dependent absolute value of the skyrmion number $|N_\text{T}|$ during a $2\pi$ rotation of the edge field for different frequencies of the edge rotation. (a) is a FM and (b) is an AFM. An edge field with an amplitude of $\SI{5}{T}$ is applied to a region of size $5 \times 14$ at the left racetrack edge.}
\label{fig:topoN_onerotation}
\end{figure}

\begin{figure}[hbt!]
\includegraphics[width=\linewidth]{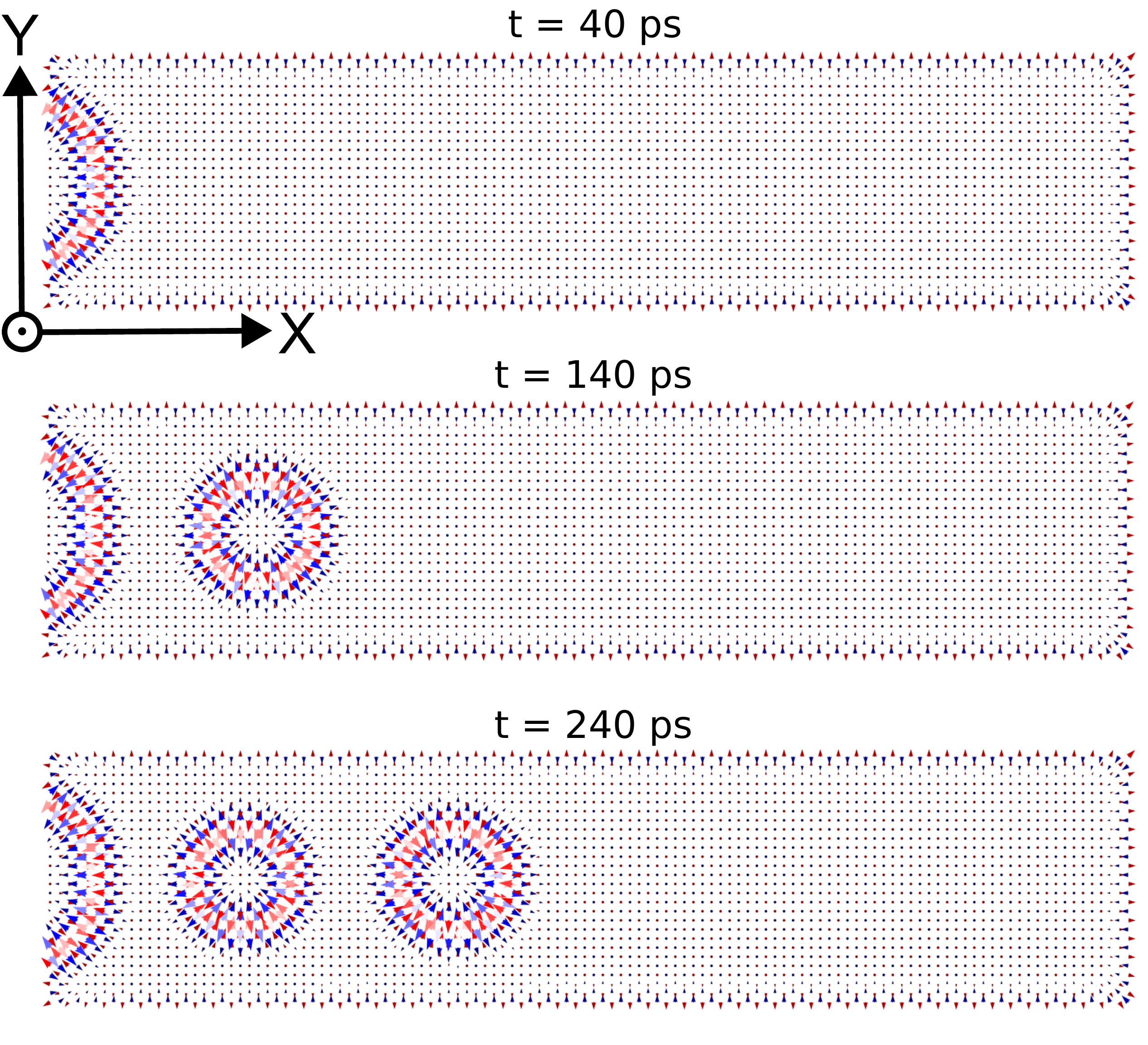}
\caption{Creation of a train of antiferromagnetic skyrmions by several 2$\pi$ cycles of a rotating edge field with a frequency $\nu=\SI{10}{GHz}$. The amplitude of the edge field is 10 T and the widths of the edge field area are given by $N_x = 5$ and $N_y = 14$.}
\label{fig:afm_skyr_rotations}
\end{figure}

To analyse the skyrmion creation process more precisely, we plotted the skyrmion number $N_{\text{T}}$ against the rotation angle of the effective edge field $\phi$, as shown in Fig.~\ref{fig:topoN_onerotation}. For a ferromagnetic racetrack, the rotation frequencies range from \SI{10}{MHz} to \SI{10}{GHz}, with an amplitude of the localized rotating effective field $B_{\text{edge}} = \SI{5}{T}$. In accordance with \cite{Schaeffer:SciRep2020}, the skyrmion number changes smoothly due to the non-homogeneous magnetization at the boundary of the racetrack, with a full skyrmion ($|N_\text{T}|=1$) appearing at a rotation angle of $\phi\approx 270^\circ$. The creation of ferromagnetic skyrmions is prohibited for frequencies exceeding \SI{10}{GHz}. Based on this result, we conclude that the relaxation time of ferromagnetic skyrmions is on the order of a few nanoseconds. Remarkably, antiferromagnetic skyrmions can be reliably produced at frequencies exceeding \SI{100}{GHz}. Even at \SI{100}{GHz}, where the shape of the function $|N_\text{T}(\phi)|$ has changed slightly, the antiferromagnetic skyrmions were created successfully. Hence, according to our simulations, the relaxation times of the antiferromagnetic skyrmions are on the order of picoseconds. Similarly to the ferromagnetic skyrmions \cite{Siegl:PhysRevB2022,Schaeffer:SciRep2020}, repeated application of the rotated edge field leads to the creation of trains of antiferromagnetic skyrmions, as seen in Fig.~\ref{fig:afm_skyr_rotations} and supplementary video \cite{SuppV}, that can be used for spintronic applications.

\section{\label{sec:results}Creation of antiferromagnetic skyrmions by switching edge fields}

So far, we have demonstrated the ability to create and manipulate trains of ferromagnetic or antiferromagnetic quasiparticles via local rotating edge fields, in a controlled manner and without the need for currents or global driving fields. Due to the directional nature of the DMI and the repulsion between the skyrmions, creating an additional object causes the existing quasiparticles to move in a given direction at a well-defined velocity. The use of rotating local fields is a state-of-the-art procedure in toggle MRAM applications \cite{Wang:IEEE2019}. However, these fields are weaker than the ones discussed above. To enhance the functionality of the proposed current-free racetrack, we are exploring the potential of using non-rotating local effective fields $\mathbf{B}_{\rm edge}$ to switch the staggered magnetization between two stable orientations and create a topological particle. This should be feasible given the chiral nature of the DMI, and it would greatly simplify the writing process. In the case of FMs, $\mathbf{B}_{\rm edge}$ may correspond to local magnetic fields. In the case of AFMs, however, a different mechanism must be used, %corresponds to any force acting on the edge; 
for example, an exchange interaction with a neighboring FM.
\begin{figure}[hbt!]
\includegraphics[width=\linewidth]{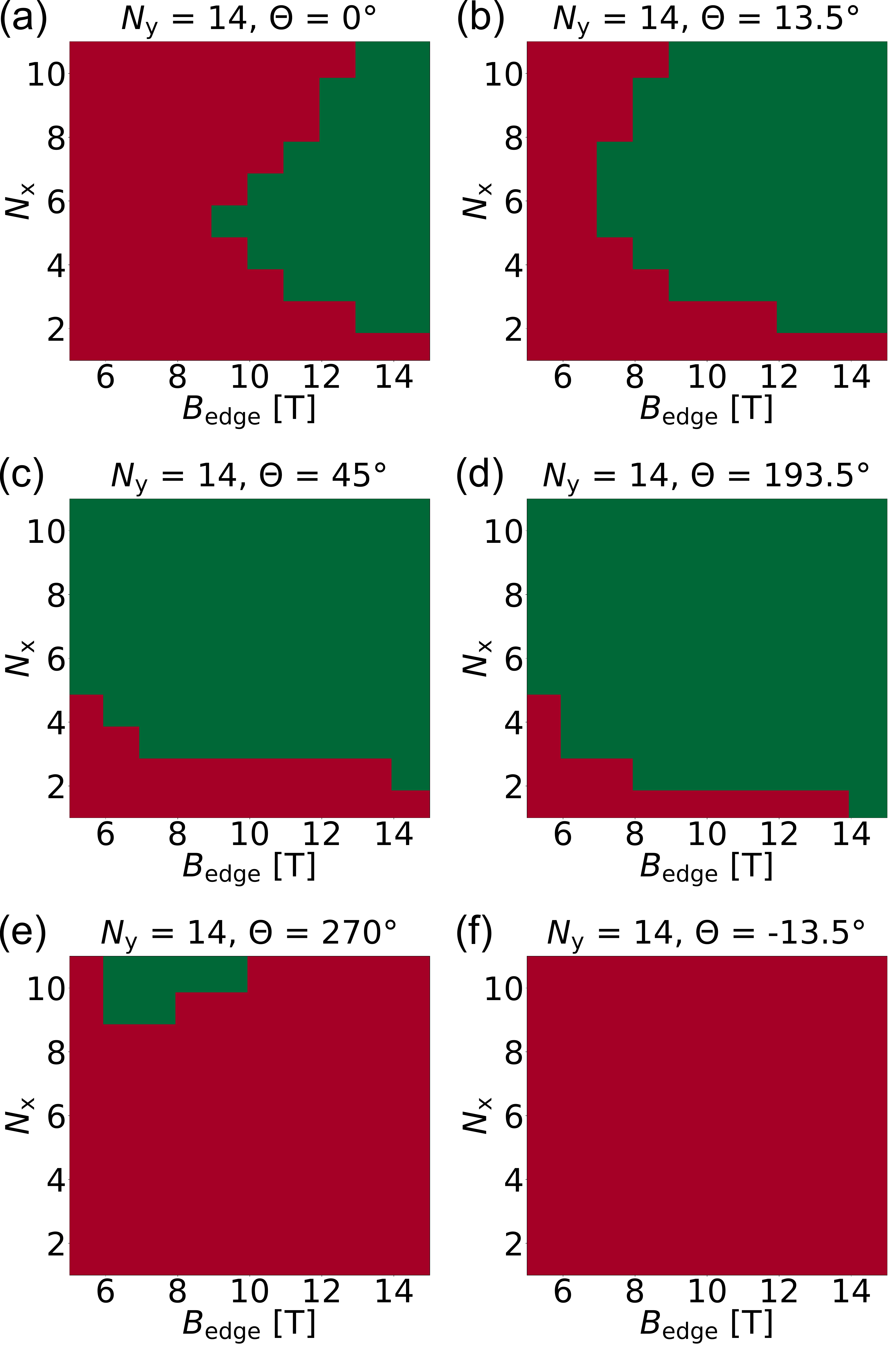}
\caption{Antiferromagnetic skyrmion creation diagrams using switching edge fields for different pairs of edge field amplitudes $B_\text{edge}$ and numbers of rotated edge rows $N_x$. The angle $\theta$ of the edge field with respect to the $z$ axis is: (a) $\theta=0$°, (b) $\theta=13.5$°, (c) $\theta=45$°, (d) $\theta=193.5$°, (e) $\theta=270$°, and (f) $\theta=-13.5$°. If the direction of the edge field is given by a vector $\hat{\mathbf{e}}_\theta$ the field switches periodically between $B_\text{edge}\hat{\mathbf{e}}_\theta$ and $-B_\text{edge}\hat{\mathbf{e}}_\theta$.}
\label{fig:afm_switching}
\end{figure}

Fig.~\ref{fig:afm_switching} shows phase diagrams that map the successful creation of skyrmions in the antiferromagnetic racetrack for various orientations of the effective edge field $B_{\rm edge}^{\rm eff}$ with respect to the vertical axis and for different $N_y$. Given a certain orientation, the sign of the edge field is flipped periodically. As can be seen in Fig.~\ref{fig:afm_switching} (c) and (d), there are optimal angles, which we call 'sweet angles', that lead to the successful creation of skyrmions across the entire range of field strengths and addressed edge rows. Of particular interest is the fact that a small adjustment to the effective edge field can suppress skyrmion formation (see Fig.~\ref{fig:afm_switching} (f)), or conversely, lead to successful events. Particularly, the negative $\Theta$ suppresses the formation of antiferromagnetic skyrmions completely. The following section discusses the effectiveness of the proposed procedure and compares the directionality of successful edge rotation in FM and AFM cases.

\begin{figure}[t!]
\includegraphics[width=\linewidth]{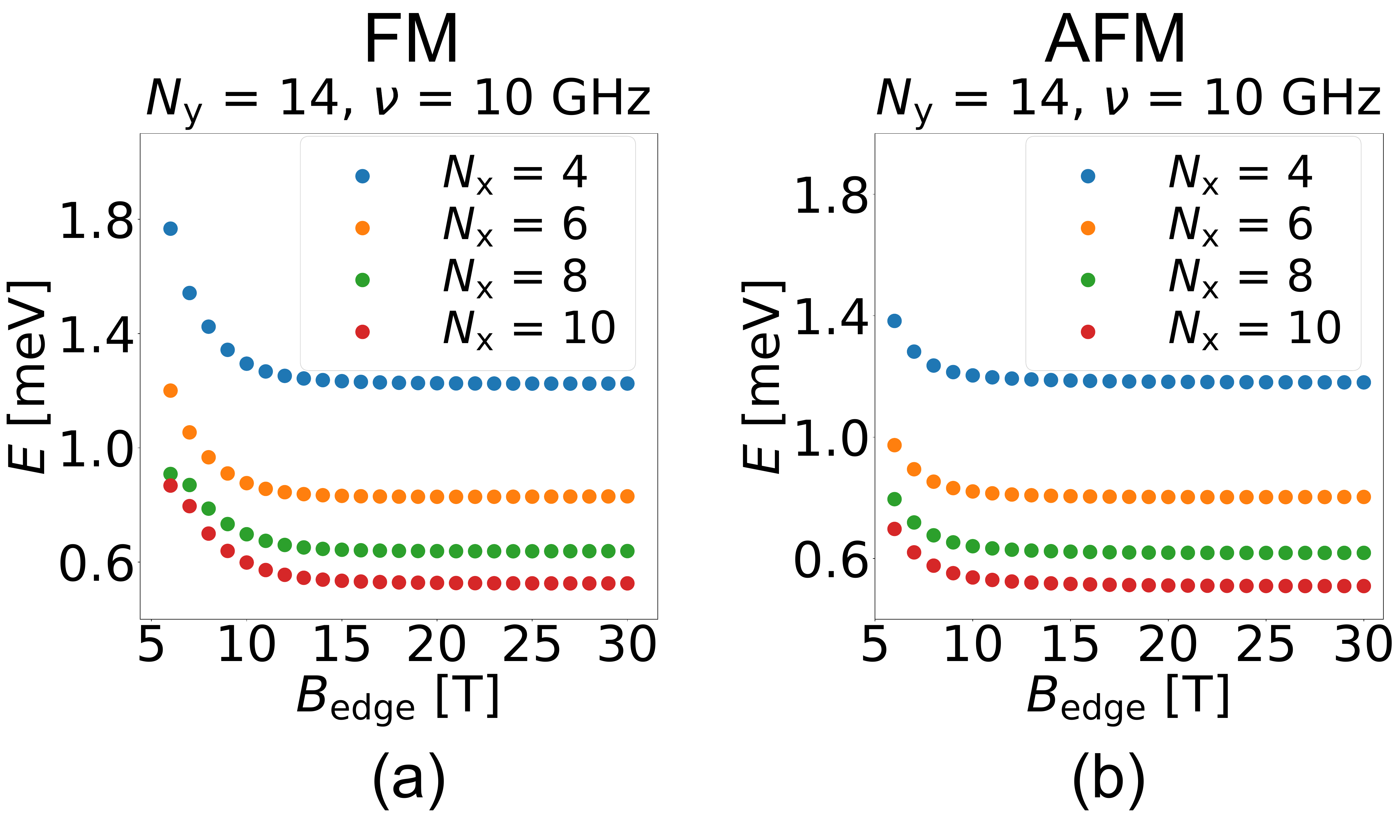}
\caption{Estimation of the work per spin performed by a rotating edge field to successfully create a skyrmion, depending on the amplitude of the edge field $B_\text{edge}$. The edge field performed a $2\pi$ rotation with a frequency $\nu=\SI{10}{GHz}$. The trajectories are given for different numbers of rotated edge rows $N_x$. (a) shows data for ferromagnetic skyrmions and (b) antiferromagnetic ones.}
\label{fig:energieestimate}
\end{figure}

\section{\label{sec:theory} Sense and efficiency of the edge rotation}

As previous publications have shown, the successful formation of quasiparticle trains is only possible if the direction of edge rotation coincides with that of the DMI. If the edge rotation is opposite to that of the DMI, the quasiparticles can be removed from the racetrack (see Figs. S1, S2 in the supplementary information to \cite{Schaeffer:SciRep2020}). This raises the question whether this is still the case for antiferromagnetic skyrmions. To address this, we remark that we can map an antiferromagnetic system to a ferromagnetic system by applying a gauge transformation depending on the sublattice type to which the spin vector $\mathbf{S}_i$ belongs. Let $A$ and $B$ be the sets of all lattice sites belonging to the sublattice types A and B, respectively. Then we define the gauge transformed spins $\tilde{\mathbf{S}}_i$ as
\begin{equation}
\tilde{\mathbf{S}}_i = 
\begin{cases}
    \mathbf{S}_i, & \text{if } i \in A \\
    -\mathbf{S}_i, & \text{if } i \in B
\end{cases}.
\end{equation}
Since we only sum the square of the z-components in the magnetic anisotropy term $H_\text{MAE}=K \sum_i \left( \mathbf{S}_i^z \right)^2$, it becomes evident that this term remains invariant under the gauge transformation.
Let $i \in A$ be an arbitrary lattice site from the sublattice A. Then we define $B_i$ as the set of all nearest neighbor sites on sublattice B with regards to the site $i$. This transforms the summation in Eq.~\eqref{eq:hamiltonian} as $\sum_{<ij>} \leftrightarrow \sum_{\substack{i \in A \\ j \in B_i}}$.  With this, the gauge transformed Hamiltonian $\tilde{H}$ reads
\begin{align}
\tilde{H} &= \sum_{\substack{i \in A \\ j \in B_i}} \left( -J \right) \tilde{\mathbf{S}}_i \cdot \tilde{\mathbf{S}}_j - \mathbf{D}_{ij} \cdot \tilde{\mathbf{S}}_i \times \tilde{\mathbf{S}}_j + H_\text{MAE}, \nonumber\\
&= \sum_{\substack{i \in A \\ j \in B_i}} \left( -J \right) \mathbf{S}_i \cdot \left( -\mathbf{S}_j \right) - \mathbf{D}_{ij} \cdot \mathbf{S}_i \times \left(-\mathbf{S}_j \right) + H_\text{MAE}, \nonumber\\
&= J \sum_{<ij>} \mathbf{S}_i \cdot \mathbf{S}_j + \sum_{<ij>} \mathbf{D}_{ij} \cdot \mathbf{S}_i \times \mathbf{S}_j - K \sum_i \left( \mathbf{S}_i^z \right)^2,
\label{eq:hamiltonian_gauge}
\end{align}
which is exactly the definition of the atomistic Hamiltonian Eq.~\eqref{eq:hamiltonian} in the non-transformed spin variables $\mathbf{S}_i$, however, with negative sign in the exchange and DMI constant. Therefore, an antiferromagnetic spin system described by the spin variables $\mathbf{S}_i$ can be mapped onto a ferromagnetic system in the spin variables $\tilde{\mathbf{S}}_i$ with opposite DMI-vectors. Consequently, for $D_x/D_y>0$ a counterclockwise rotation of the edge field is required to create a skyrmion in an AFM. This is opposite to a ferromagnetic system, where $D_x/D_y>0$ demands a clockwise rotation for successful creation of a skyrmion according to \cite{Siegl:PhysRevB2022,Schaeffer:SciRep2020}.

The following aims at estimating the potential energy costs of creating skyrmions using the edge field rotation technique.
To do this, we first define our effective local field, which is the sum of all interactions
\begin{equation}
\mathbf{B}_{\text{eff},i} = -\frac{1}{\hbar \gamma} \frac{\partial H}{\partial \mathbf{S}_i}.
\end{equation}
With this we can write the total differential of the energy as
\begin{equation}
\dd H_i = \frac{\partial H}{\partial \mathbf{S}_i} \cdot \dd \mathbf{S}_i = - \hbar \gamma \mathbf{B}_{\text{eff},i} \cdot \dd \mathbf{S}_i.
\label{eq:diff_hamiltonian}
\end{equation}
Since we are only interested in the work performed by the external rotating edge field, we will regard only its contribution and replace $\mathbf{B}_{\text{eff},i} \rightarrow \mathbf{B}_{\text{edge},i}$ in the next steps. This work can then be expressed using the total derivative $\mathbf{B}_{\text{edge},i} \cdot \dd \mathbf{S}_i = \dd\left( \mathbf{B}_{\text{edge},i} \cdot \mathbf{S}_i \right) - \mathbf{S}_i \cdot \dd\mathbf{B}_{\text{edge},i}$ of Eq. \eqref{eq:diff_hamiltonian}. Here, the second term of the right-hand side of the equation represents the amount of work required to make an infinitesimal change to the edge field for a given configuration of magnetic moments $\mathbf{S}_i$. Since we assume adiabatic change of the magnetic configuration, $\dd\left( \mathbf{B}_{\text{edge},i} \cdot \mathbf{S}_i \right) = 0$. Consequently, the estimated amount of work $\dd W_{{\rm edge},i}$ to change the edge field by $\dd \mathbf{B}_{\text{edge},i}$ reads
\begin{equation}
    \dd W_{{\rm edge},i} = \hbar \gamma \mathbf{B}_{\text{edge},i} \cdot \dd \mathbf{S}_i
    \label{eq:diff_hamiltonian_ext}
\end{equation}
Now, one only has to sum Eq. \eqref{eq:diff_hamiltonian_ext} over all magnetic moments for which the edge field is non-zero and integrate over one period $T$ of the rotation to obtain a formula for the required energy to create a skyrmion. We will denote this energy as $\Delta H_{\text{skyr}}$. This already assumes that the magnitude of the edge field as well as the number of rotated edge rows have been chosen in accordance with the stability diagrams discussed previously, see Fig.~\ref{fig:creation_diagrams}.

In the limit of infinitesimal small time steps $\dd t$ 
\begin{equation}
\Delta H_{\text{skyr}} =  \hbar \gamma \sum_i \int_T \mathbf{B}_{\text{edge},i} \cdot \frac{\partial \mathbf{S}_i}{\partial t} \dd t,
\end{equation}
with the sum running over the lattice sites $i$ involved in the edge rotation, can be approximated numerically by a finite time grid. 
% \begin{equation}
% \Delta H_{\text{skyr}} \approx  \hbar \gamma \sum_i \sum_{j=1}^N \mathbf{B}_{\text{edge,j}} \cdot \dd \textbf{S}_j,
% \label{eq:creation_energy}
% \end{equation}
% where $N$ denotes the number of simulation steps for one full rotation of the edge field. Furthermore, we use the following notations: $\mathbf{B}_{\text{edge,j}} = \mathbf{B}_{\text{edge}}(t=j \cdot \dd t)$ and $\dd \textbf{S}_j = \textbf{S}(t=j \cdot \dd t) - \textbf{S}(t=(j-1) \cdot \dd t)$.
Using the procedure described above, we estimate that creating one skyrmion for the material parameters used in this paper and in a field of \SI{10}{T} requires a work of $\Delta H_{\rm skyr} = \SI{75}{meV}$. Meanwhile, the energy difference between the ferromagnetic racetrack without a skyrmion and the racetrack with a skyrmion is $H_{\rm skyr}-H_{\rm FM} = \SI{62}{meV}$. Hence, the proposed procedure shows a fairly high efficiency in an ideal case. As can be seen in Fig.~\ref{fig:energieestimate}, a comparison between ferromagnetic and antiferromagnetic skyrmion creation shows only minor differences. When considering large edge fields, the energies seem to be the same for FM and AFM systems. For relatively small edge fields the antiferromagnetic skyrmions are more efficient.

\section{\label{sec:conclusion}Conclusion}

In conclusion, as an extension of the setup for the controlled creation, deletion, and propulsion of ferromagnetic quasiparticles in racetracks, we have introduced the theoretical concept of a setup that allows the same functionality for antiferromagnetic skyrmions. Additionally, we demonstrated that the rotated field could be successfully replaced by switching the edge magnetization between two collinear states. This procedure makes the creation and processing of topological particles by local excitations, rather than global fields or currents, more robust against positional fluctuations, thus simplifying its technical realization. Our investigation has also revealed several interesting fundamental aspects of the antiferromagnetic skyrmions. In particular, antiferromagnetic skyrmions require a different rotation direction compared to their ferromagnetic counterparts, and they can be created by local excitation at frequencies that are orders of magnitude larger than those for ferromagnetic skyrmions. In this study, we assume that the system is well below the critical temperature at which skyrmions are stable. In this case, skyrmion creation by edge rotations can be realized successfully. Further studies are needed to investigate the stability and formation of skyrmions by edge rotation at non-zero temperatures and for different Gilbert dampings.

\section*{Supplementary Material}
The supplementary material includes a video of a continuous creation of antiferromagnetic skyrmions by rotation of the edge field.
\section*{Acknowledgments}
Authors acknowledge financial support provided by the Deutsche Forschungsgemeinschaft (DFG) via Project No. 514141286.

\bibliography{references}

\end{document}